\documentstyle [12pt] {article}

\catcode`@=12
\begin{document}
\thispagestyle{empty}
\begin{flushright} UCRHEP-T166\\June 1996\
\end{flushright}
\vspace{0.5in}
\begin{center}
{\large	\bf Framework for Naturally Light Singlet Neutrinos\\}
\vspace{1.8in}
{\bf Ernest Ma\\}
\vspace{0.3in}
{\sl Department of Physics\\}
{\sl University of California\\}
{\sl Riverside, California 92521, USA\\}
\vspace{1.5in}
\end{center}
\begin{abstract}\
The totality of present neutrino data seems to require four light neutrinos, 
but only three of them can be the neutral components of left-handed lepton 
doublets.  To accommodate one or more naturally light singlet neutrino(s), 
an extra U(1) gauge factor is proposed to implement an analogous seesaw 
mechanism which accounts for the light doublet neutrinos.  Using the 
constraints of anomaly cancellation, the property of this U(1) is determined. 
The most attractive theoretical framework is that of a supersymmetric 
$SU(3)_C \times SU(2)_L \times U(1)_Y \times U(1)_N$ model already proposed.
\end{abstract}

\newpage
\baselineskip 24pt

There are at present a number of neutrino experiments with data\cite{1,2,3} 
which can be interpreted as being due to neutrino oscillations.\cite{4}  
Solar data\cite{1} indicate the oscillation of neutrinos differing in the 
square of their masses of the order $\Delta m^2 \sim 10^{-5} {\rm eV}^2$ 
for the matter-enhanced solution\cite{5} or $\Delta m^2 \sim 10^{-10} 
{\rm eV}^2$ for the vacuum solution.  Atmospheric data\cite{2} indicate 
possible oscillation of $\Delta m^2 \sim 10^{-2} {\rm eV}^2$.  More 
recently, the LSND (Liquid Scintillator Neutrino Detector) experiment has 
obtained results\cite{3} which indicate possible oscillation of 
$\Delta m^2 \sim 1~{\rm eV}^2$.  To accommodate all the above data as 
being due to neutrino oscillations, it is clear that four neutrinos are 
needed to have three unequal mass differences.  Since the invisible width of 
the $Z$ boson is already saturated with the three known doublet neutrinos 
$\nu_e$, $\nu_\mu$, and $\nu_\tau$, {\it i.e.} from $Z \rightarrow 
\nu \bar \nu$, one must then have a fourth neutrino 
which does not couple to the $Z$ boson, {\it i.e.} a singlet.
The question is why such a singlet neutrino should be light.

Let us review our current understanding of why the three known doublet 
neutrinos are light.  It is called the seesaw mechanism\cite{6}.  First, 
we assume that for each left-handed doublet neutrino $\nu$, there is a 
right-handed singlet neutral fermion $N$ which couples to the former 
through the usual Higgs scalar doublet $\Phi = (\phi^+, \phi^0)$ of the 
standard model.  The mass matrix spanning $\nu$ and $N$ is then given by
\begin{equation}
{\cal M} = \left[ \begin{array} {c@{\quad}c} 0 & m_D \\ m_D & m_N \end{array} 
\right], 
\end{equation}
where $m_D$ comes from the vacuum expectation value of $\phi^0$ and $m_N$ is 
a Majorana mass allowed by the fact that $N$ is a singlet under the 
$SU(3)_C \times SU(2)_L \times U(1)_Y$ gauge symmetry of the standard model. 
The origin of $m_N$ is presumably from new physics at a much higher energy 
scale, so it should be large.  The zero in the (11) entry of $\cal M$ is 
protected by the standard gauge group and the fact that there is no scalar 
triplet.  As a result, the neutrino mass is given by the well-known formula
\begin{equation}
m_\nu \sim {m_D^2 \over m_N}.
\end{equation}
For a given value of $m_D$, say of the order of the corresponding 
charged-lepton mass, $m_\nu$ can be very small for a very large $m_N$.

To have a very light singlet neutrino $S$, let us make sure that its mass 
is zero by itself, just as $m_\nu$ would be without $N$.  To give this zero 
the analogous level of protection as the (11) entry of Eq.~(1), assume an 
extra U(1) gauge factor, under which $N$ is trivial, but not $S$.  Hence 
$m_N$ is still allowed, but not $m_S$.  We now let $S$ couple to $N$ through 
a singlet scalar boson $\chi$ which develops a nonzero vacuum expectation 
value, thereby breaking this extra U(1).  The mass matrix spanning $\nu$, 
$N$, and $S$ is then given by
\begin{equation}
{\cal M} = \left[ \begin{array} {c@{\quad}c@{\quad}c} 0 & m_D & 0 \\ 
m_D & m_N & m_X \\ 0 & m_X & 0 \end{array} \right].
\end{equation}
In the limit of large $m_N$, this reduces to
\begin{equation}
{\cal M} = \left[ \begin{array} {c@{\quad}c} m_D^2/m_N & m_D m_X/m_N \\ 
m_D m_X/m_N & m_X^2/m_N \end{array} \right]
\end{equation}
for the two light neutrinos, with mass eigenvalues 0 and $(m_D^2 + m_X^2)
/m_N$, corresponding to the states $\nu \cos \theta - S \sin \theta$ and 
$\nu \sin \theta + S \cos \theta$ respectively, where $\tan \theta = 
m_D/m_X$.

So far, we have not specified how the usual quarks and leptons transform 
under this new $U(1)'$.  However, since they must acquire masses by coupling 
to scalar doublets, the following Yukawa interaction terms must exist:
\begin{eqnarray}
(\nu_e, e) e^c (\phi_1^0, \phi_1^-), &~& (\nu_e, e) N (\phi_2^+, \phi_2^0), \\ 
(u,d) d^c (\phi_3^0, \phi_3^-), &~& (u,d) u^c (\phi_4^+, \phi_4^0),
\end{eqnarray}
where we have adopted the notation that all fermions are left-handed and the 
superscript $c$ denotes the charge-conjugated state.  We have assumed four 
different scalar doublets with $U(1)'$ assignments $N_{1,2,3,4}$ 
respectively.  We also assign $N_S$ to $S$ (and thus $-N_S$ to $\chi$) as 
well as $N_q$ to $(u,d)$.  We now require this extended gauge model to be 
free of triangle anomalies\cite{7}.  We find
\begin{eqnarray}
Tr (T_3^2 N) = 0 &\Rightarrow& N_2 = 3 N_q, \\ Tr (Y^2 N) = 0 &\Rightarrow& 
3 N_1 + N_3 + 4 N_4 = 0, \\ Tr (Y N^2) = 0 &\Rightarrow& (N_3 + N_4) 
(6 N_2 + 5 N_3 - N_4) = 0, \\ Tr (N^3) = 0 &\Rightarrow& \sum N_S^3 = 
35 n_f \left( {{N_3 + N_4} \over 6} \right)^3,
\end{eqnarray}
where $n_f$ is the number of families and we have allowed more than just one 
$S$.

The simplest solution to Eq.~(10) is to assume that for each family, there 
are two $S$ fermions with
\begin{equation}
N_{S_1} = {1 \over 2} (N_3 + N_4), ~~~ N_{S_2} = {1 \over 3} (N_3 + N_4).
\end{equation}
Furthermore, we can choose $N_3 + N_4 = 6$ without loss of generality 
because that can be absorbed into a redefinition of the $U(1)'$ coupling. 
We now consider Eqs.~(8) and (9) to obtain a family of solutions:
\begin{equation}
N_1 = -2-N_4, ~~~ N_2 = N_4-5, ~~~ N_3 = 6-N_4, ~~~ N_{S_1} = 3, ~~~ 
N_{S_2} = 2.
\end{equation}
Note that there is no solution for less than four different doublets, 
{\it i.e.} $N_1 = N_3$ or $N_1 = -N_2$, {\it etc.} are impossible because 
they always lead to $N_3 + N_4 = 0$.  Note also that if $N$ is a solution 
to Eqs.~(7) to (10), then $aN+bY$ is also a solution; hence a single-parameter 
family of solutions is the best we can do without further assumptions.

If we allow $N_3 + N_4 = 0$, then Eqs.~(8) and (9) can be satisfied with 
$N_1 = -N_2 = N_3 = -N_4$, {\it i.e.} we need only one scalar doublet. 
On the other hand, Eq.~(10) must now be satisfied by either pairs of singlets 
with opposite $N_S$, which would then form Dirac masses with each other, 
contrary to the aim of this investigation, or by having a large number of 
singlets with just the right $N_S$ assignments.  An example of the latter 
is to have one singlet with $N_S = 3$, three others with $N_S = -2$, and 
yet three more with $N_S = -1$.

Another way to implement the idea of a light singlet is to use a variation 
of the seesaw mechanism.  Let $S$ couple to 
two fermion doublets $(\nu_E, E)$ and $(E^c, N_E^c)$ 
with electric charge assignments $(0,-1)$ and (1,0) respectively, and assume 
also that these two doublets can combine to have a large Dirac mass.  The 
mass matrix spanning $\nu_E$, $N_E^c$, and $S$ is then given by
\begin{equation}
{\cal M} = \left[ \begin{array} {c@{\quad}c@{\quad}c} 0 & m_E & m_1 \\ 
m_E & 0 & m_2 \\ m_1 & m_2 & 0 \end{array} \right].
\end{equation}
As a result,
\begin{equation}
m_S \sim {{2 m_1 m_2} \over m_E},
\end{equation}
and the mechanism is again seesaw, but instead of the usual 
version\cite{6} where the large mass is that of a singlet, here it is that 
of a doublet.

In order to have $\nu-S$ mixing, we must now connect the two sectors. 
If we restrict ourselves to only scalar doublets, then only the 
$\nu S$, $\nu_E N$, and $N_E^c N$ terms have to be considered.  If the 
$\nu S$ term exists, then it has to be fine-tuned to be much smaller than 
$m_D^2/m_N$ and $2m_1 m_2/m_E$ to be phenomenologically viable.  Hence it 
is much better to forbid it by the assumed $U(1)'$.  Consequently, the 
combined mass matrix spanning $\nu$, $N$, $\nu_E$, $N_E^c$, and $S$ is 
given by
\begin{equation}
{\cal M} = \left[ \begin{array} {c@{\quad}c@{\quad}c@{\quad}c@{\quad}c} 
0 & m_D & 0 & 0 & 0 \\ m_D & m_N & m_3 & m_4 & 0 \\ 0 & m_3 & 0 & m_E & m_1 \\ 
0 & m_4 & m_E & 0 & m_2 \\ 0 & 0 & m_1 & m_2 & 0 \end{array} \right].
\end{equation}
In the limit of large $m_N$ and $m_E$, the effective mass 
matrix for the light neutrinos is then
\begin{equation}
{\cal M} = \left[ \begin{array} {c@{\quad}c} m_D^2/m_N & 
(m_1 m_3 + m_2 m_4) m_D/m_N m_E \\ 
(m_1 m_3 + m_2 m_4) m_D/m_N m_E & 2 m_1 m_2/m_E 
\end{array} \right]
\end{equation}
as desired.  Note that in this scenario, the $\nu-S$ mixing is naturally 
small.  Again the consideration of anomaly cancellation leads to 
Eqs.~(7) to (10), but now $m_{1,2,3,4}$ require up to four more scalar 
doublets with $N_{6,5,8,7}$ respectively.  However, as it turns out, 
the minimum solution needs only one more.  For example, let $N_4 = 8$, 
then $N_1 = -10$, $N_2 = 3$, $N_3 = -2$, $N_5 = -N_4$, $N_6 = -N_3$, 
and the only new assignment is $N_7 = -N_8 = -5$.  Of course, we still need 
to add one singlet to break the $U(1)'$.  Note also that the $\nu S$ term 
is indeed forbidden.

The above discussion shows that in order to have naturally light singlet 
neutrinos under the protection of an extra U(1), many new particles are 
required which are otherwise unmotivated.  The question is now whether there 
is another framework where such particles are not necessary.  The fact 
that the fermion doublets introduced above, {\i.e.} $(\nu_E, E)$ and 
$(E^c, N_E^c)$, have the same standard-model gauge transformations as 
the scalar doublets $(\phi_1^0, \phi_1^-)$ and $(\phi_2^+, \phi_2^0)$ is 
a strong hint that we should consider supersymmetry and identify them as 
partners.

Assume first the following supersymmetric extension of the standard model: 
in addition to the quark and lepton superfields, each family has two 
Higgs superfields as well as one $N$ and one $S$.  The anomaly-free 
conditions now have no solution.  However, if we add two color-triplet 
superfields transforming under $SU(3)_C \times SU(2)_L \times U(1)_Y \times 
U(1)'$ as follows:
\begin{equation}
h \sim (3,1,-1/3; N_h), ~~~ h^c \sim (3^*,1,1/3; N_{h^c}),
\end{equation}
then we obtain
\begin{eqnarray}
N_1 + 3 N_q = 0, ~~~ N_h + N_{h^c} = N_1 + N_2, \\ 3(N_h - N_{h^c}) = N_1 
- 3 N_2, ~~~ N_S = - N_1 - N_2.
\end{eqnarray}
A solution is now possible, {\it i.e.}
\begin{equation}
N_q = -{1 \over 3}N_1, ~~~ N_S = -N_1-N_2, ~~~ N_h = {2 \over 3}N_1, ~~~ 
N_{h^c} = {1 \over 3}N_1 + N_2.
\end{equation}
Furthermore, the above $U(1)'$ assignments fix all the possible Yukawa 
interactions among the superfields.  For example, $(\nu_E, E)(E^c, N_E^c)S$ 
is allowed but not $(\nu_E, E)(E^c, N_E^c)$, and $(\nu_e, e)(E^c, N_E^c)$ 
is allowed but not any other term which links the singlet and doublet 
neutrino sectors.  As a result, the analog of Eq.~(15) is now
\begin{equation}
{\cal M} = \left[ \begin{array} {c@{\quad}c@{\quad}c@{\quad}c@{\quad}c} 
0 & m_D & 0 & m_3 & 0 \\ m_D & m_N & 0 & 0 & 0 \\ 0 & 0 & 0 & m_E & m_1 \\ 
m_3 & 0 & m_E & 0 & m_2 \\ 0 & 0 & m_1 & m_2 & 0 \end{array} \right],
\end{equation}
which reduces to 
\begin{equation}
{\cal M} = \left[ \begin{array} {c@{\quad}c} m_D^2/m_N & m_1 m_3/m_E \\ 
m_1 m_3/m_E & 2 m_1 m_2/m_E \end{array} \right]
\end{equation}
as desired.

Let us now consider
\begin{equation}
N' = {{-5N+2(3N_2-2N_1)Y} \over {N_1+N_2}},
\end{equation}
which is a solution to Eqs.~(18) and (19) because $N$ is.  We then have 
the following $U(1)'$ assignments:
\begin{eqnarray}
(u,d), u^c, e^c \sim 1; ~~~ d^c, (\nu_e, e) \sim 2; ~~~ N \sim 0; \\ 
(\nu_E, E), h^c \sim -3; ~~~ (E^c, N_E^c), h \sim -2; ~~~ S \sim 5.
\end{eqnarray}
Hence $U(1)'$ can be embedded into $E_6$ and
\begin{equation}
N' = 6 Y_L + T_{3R} - 9 Y_R
\end{equation}
under the $SU(3)_C \times SU(2)_L \times U(1)_{Y_L} \times SU(2)_R \times 
U(1)_{Y_R}$ decomposition with the electric charge $Q = T_{3L} + Y_L + 
T_{3R} + Y_R$.  The superfields of Eqs.~(24) and (25) make up exactly the 
fundamental {\bf 27} representation of $E_6$, and the $U(1)'$ of Eq.~(26) 
is exactly what is called $U(1)_N$ in two previous papers\cite{8}.

In conclusion, if a light singlet neutrino is needed to explain data in terms 
of neutrino oscillations, the natural framework is an extra U(1) gauge 
symmetry.  If no other new fermions are added, then the requirement of 
anomaly cancellation requires two singlet neutrinos per family and four 
scalar doublets, as well as a scalar singlet which breaks $U(1)'$ 
spontaneously.  On the other hand, if supersymmetry is included, then a model 
based on $E_6$ is the natural choice.  Details of the latter have already 
been presented\cite{8}.
\vspace{0.3in}
\begin{center} {ACKNOWLEDGEMENT}
\end{center}

This work was supported in part by the U.~S. Department of Energy under 
Grant No. DE-FG03-94ER40837.

\newpage
\bibliographystyle{unsrt}

\end{document}